\documentclass[pre,reprint, showpacs]{revtex4-1}
\usepackage{amsmath, graphicx}
\begin{document}
\title{The effects of polydispersity and metastability on crystal growth kinetics}

\date{\today}

\author{J.~J.~Williamson}
\email{pyjjw@leeds.ac.uk}
\author{R.~M.~L.~Evans}
\affiliation{Soft Matter Group, School of Physics and Astronomy, University of Leeds, Leeds LS2 9JT United Kingdom}

\begin{abstract} 
We investigate the effect of metastable gas-liquid (G-L) separation on crystal growth in a system of either monodisperse or slightly size-polydisperse square well particles, using a simulation setup that allows us to focus on the growth of a \textit{single} crystal. Consistent with experiment and theoretical free energy considerations, we find that, inside the metastable binodal, a layer of the gas phase `coats' the crystal as it grows. Crucially, the effect of this metastable G-L separation on the crystal growth rate is qualitatively altered by a very small degree of polydispersity as compared to the monodisperse case, suggesting a highly local fractionation process which is facilitated by the gas layer. Our results show that polydispersity and metastability, both ubiquitous in soft matter, must be considered in tandem if their dynamical effects are to be understood.
\end{abstract}


\maketitle

\section{\label{sec:intro}Introduction}

Substances in the category of `soft condensed matter' (colloidal suspensions, polymers, proteins etc.), as well as having widespread industrial and medical importance, are interesting in their own right because they exhibit physics analogous to that of simpler molecular or atomic systems, but on much longer time- and length-scales \cite{Prasad2007}. The analogy stems from the importance of thermal fluctuations in the microscopic dynamics of the constituents, which means that the same thermodynamic and statistical mechanical approaches can be applied in both cases \cite{Onsager1949}. Nonetheless, the study and exploitation of soft matter presents a number of challenges, two of which are the focus of the present work. 

Firstly, a collection of mesoscopic particles which are nominally the same will almost certainly be \textit{polydisperse}, i.e.\  will exhibit a distribution in properties such as size, charge, shape or chemical makeup. This is in contrast to e.g.\ a molecular fluid of pure water in which, in a strict sense, every molecule is identical. Efforts by various workers to describe the thermodynamics of polydisperse systems \cite{Evans1998, Evans2000, Sollich2009, Sollich2011, Fasolo2004, Wilding2004, Fantoni2006, Bartlett1999} have resulted so far in a reasonably firm understanding of the effects of a mild degree of polydispersity on the \textit{equilibrium} phase diagrams of hard-spheres and related systems. Beyond this, polydispersity remains a pervasive but poorly understood complicating factor in soft matter physics -- in particular, its effects on phase ordering kinetics as actually enacted in real systems are unclear \cite{Zaccarelli2009, Williams2008, Martin2003, Martin2005, Schope2007, Liddle2011}, and quantitative theoretical work thereon is almost non-existent \cite{Warren1999, Evans2001}.

Secondly, and in common with simpler systems, soft matter's path towards equilibrium may be influenced by the presence of \textit{metastable} states. These are the result of local minima in the free energy landscape which, although they do not fully minimise the free energy of the system (and therefore do not appear in the equilibrium phase diagram), may be encountered as an intermediate stage. They may be long-lived, especially if the system must overcome some free energy barrier in order to reach its global minimum. A prime example is the metastable gas-liquid (G-L) separation of attractive particles, which may subsequently progress to equilibrium crystal-fluid coexistence \cite{Liddle2011}. The influence of metastability on phase ordering is a problem of general importance, with particular application in e.g.\ colloid-polymer mixtures and protein crystallisation \cite{Anderson2002, Dixit2000, Haas2000, Haas1999}. 

In this paper, we perform simulations which focus on crystal growth (as distinct from nucleation) in a model colloidal system exhibiting the metastable G-L coexistence described above. By varying the interaction strength and size-polydispersity, we study the effects of metastability and polydispersity on the crystal growth dynamics. For our parameters, we find that the metastable G-L separation results in a gaseous layer coating the growing crystal, an observation which we relate to theoretical free energy curves and existing experimental work. When polydispersity is present, this layer can act to speed up crystal growth, in contrast to the monodisperse case in which crystal growth is slowed by the presence of the gas layer. Interpreted as a dynamical phenomenon, this suggests that fractionation at the crystal interface, relying on self diffusion, is facilitated by the gas layer. Our results shed light on the influence of polydispersity and metastability on the crystal growth process, demonstrating the complex way in which these factors can interact. 

The structure of the paper is as follows. In Section~\ref{sec:simulation} we describe the simulation apparatus and the parameters used, with reference to the equilibrium monodisperse phase diagram of our system. In Section~\ref{sec:theory} we outline two pertinent theoretical aspects of the work, viz. the free energy landscape of the system, and the diffusive growth of split interfaces of the kind exhibited in our simulations. In Section~\ref{sec:results} we present the simulation results and discuss the effects of metastability, polydispersity, and the two combined on the crystal growth process. Finally, in Section~\ref{sec:fractionation}, we outline and briefly test a possible explanation for the complex interaction between metastability and polydispersity, in the course of which we measure fractionation (de-mixing) associated with crystal formation, and present novel findings relating to local size correlations in polydisperse systems. We conclude and motivate future work in Section~\ref{sec:conclusions}.

\section{\label{sec:simulation}Simulation}

\subsection{\label{sec:kmc}Kinetic Monte Carlo algorithm}

The simulation is built on a Kinetic Monte Carlo (KMC) algorithm, in which the available trial moves are limited to small stochastic `hops', representing the inertia-free Brownian motion of colloidal particles dispersed in a solvent. Therefore, in contrast to equilibrium MC methods, KMC is suitable for studying the dynamics of phase transitions since unphysical moves such as cluster rearrangements, particle resizing etc., which are helpful in speeding up equilibration, are not used. In common with comparable Molecular or Brownian Dynamics methods, hydrodynamic interactions are neglected. Further details of our implementation are available in Ref.~\cite{Williamson2012}.

\subsection{\label{sec:geometry}Parameters, crystal template}

The system studied consists of $N = 5000$ spherical particles with diameters $d_{n}$ drawn from a Bates (pseudo-Gaussian) distribution of polydispersity (defined as the ratio of the standard deviation to the mean) $\sigma = 0$ (`monodisperse'), $\sigma = 0.03$ or $\sigma = 0.06$ \footnote{The Bates distribution is a sum of \protect{$n$} random numbers uniformly distributed on the unit interval -- in our implementation \protect{$n = 4$}.}. The mean hard core diameter $\langle d \rangle$ sets the length unit, and the time unit is the mean time $t_{d}$ taken for a free particle of diameter $\langle d \rangle$ to diffuse a distance equal to its own diameter. The hard particle cores are surrounded by pairwise additive square wells of depth $u$ and range $\lambda = 1.15$. The interaction range between specific particles $i$ and $j$ depends on the particle diameters in a scalable fashion \cite{Williamson2012, Evans2001}:

{\begin{equation}
V(r) = 
\begin{cases}
\infty & \text{if } r\leq d_{ij} \\
-u & \text{if } d_{ij} < r\leq \lambda d_{ij}\\
0 & \text{if } r > \lambda d_{ij}
\end{cases}
\label{eqn:scalablesquarewell}
\end{equation}}

\noindent with $d_{ij} = (d_{i} + d_{j})/2$. While the range parameter $\lambda$ is the same for all particles, Equation~\ref{eqn:scalablesquarewell} shows that the \textit{actual} range between a given particle pair depends multiplicatively on the size of their hard cores. 

\begin{figure}
\centering
\includegraphics[width=8.6cm]{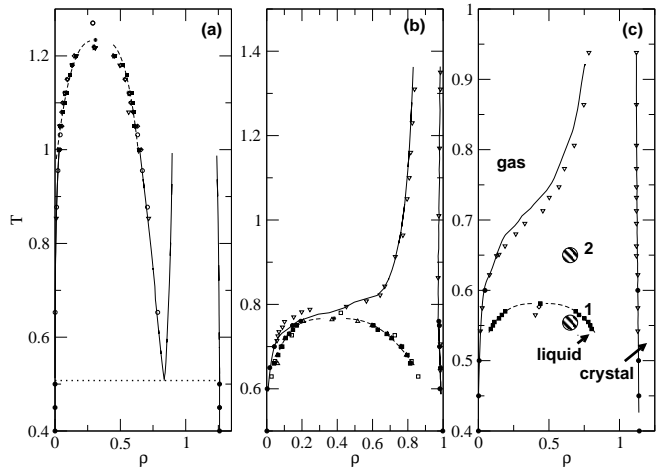}
\caption{\label{LGK}Phase diagrams reproduced with permission from Ref.~\cite{Liu2005} for (a) $\lambda = 1.5$, (b) $\lambda = 1.25$ and (c) $\lambda = 1.15$ in the density-temperature plane. With our choice of units, $T = T_{\textrm{eff}} \equiv 1/u$ and $\rho = 6\phi / \pi$. As $\lambda$ decreases, the gas-liquid coexistence becomes metastable with respect to crystal-gas separation. The relevant diagram for our system is (c). The points corresponding to the $\phi, u$ coordinates used in our study are indicated with hashed circles: point 1 is `inside the G-L binodal' or `gas-mediated' and point 2 is `outside the binodal' or `gas-free'. The crystal, the (stable) gas, and the liquid side of the metastable G-L binodal are marked.}
\end{figure}

The parent volume fraction (hereafter also referred to as `concentration') is $\phi_{p} = 0.34$. In the monodisperse limit, the values of $\phi_{p}$ and $\lambda$ are such that the equilibrium state is a coexistence of crystal and gas \cite{Liu2005}. However, by choosing $u$ (and therefore the effective temperature $T_{\textrm{eff}} = 1 / u$) appropriately, the system's state point can be made to lie either inside or outside the metastable gas-liquid binodal, as shown in FIG.~\ref{LGK} (reproduced from Ref.~\cite{Liu2005}). Given that the critical effective temperature corresponds to $u = 1.72$, we use a well depth of $u = 1.82$ inside the G-L binodal and $u = 1.54$ outside, where the energy unit is $k_{B}T$. In soft matter, one is often free to vary the interaction strength in this way, for instance by adding more or less polymer into a colloid-polymer mixture. 

Since our aim is to study the process of crystal \textit{growth}, not nucleation, we introduce crystallisation artificially by placing a $10 \times 10$ template of immobilised particles, arranged in the (100) face of the FCC lattice, at one end of the simulation cell. The simulation cell is long in the $x$ axis ($L_{x} \approx 66$), and relatively short in the (periodic) $y$ and $z$ dimensions ($L_{y} = L_{z} \approx 11$), so that the growth of the templated crystal along the $x$ axis can be studied for a long period of time. The template's lattice parameter is set to match the volume fraction of the equilibrium crystal \cite{Liu2005}. Increasing the $y$ and $z$ dimensions of the lattice had no measurable effect on the results presented. Crystalline particles are identified by constructing bond order vectors $\textbf{q}_{6}$ from the spherical harmonics $Y_{6m}$ \cite{tenWolde1996, Williams2008}: Particle $i$ is flagged crystalline if $\sum_{j=1}^{N_{b}(i)}\textbf{q}_{6}(i) \cdot \textbf{q}_{6}(j) > 8.5$, where the sum is over the $N_{b}(i)$ neighbours within $\sqrt{1.45r_{0}^2}$ of particle $i$, $r_{0}$ being the average separation between particle $i$ and its closest 6 neighbours. The result of the crystal identification algorithm can be seen in FIG.~\ref{flagging}.

\begin{figure}
\centering
\includegraphics[width=8.6cm]{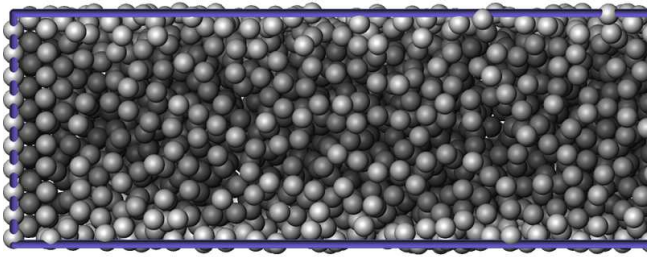}
\caption{\label{template}Illustrative snapshot of a monodisperse system with $u=1.54$ at $t \approx 100$, showing the crystal templating method. The FCC template is positioned at $x = 0$, and in this case 1 or 2 further crystalline layers have so far been deposited from the bulk fluid. Approximately half the length of the simulation cell is shown. Simulation snapshots were produced using OVITO \cite{Ovito}.}
\end{figure}

After initialising the system as an amorphous fluid, the square well attraction is turned on and the template positioned, defining $t = 0$. An illustrative simulation snapshot shortly thereafter is shown in FIG.~\ref{template}. In principle, the nucleation of independent crystals in the bulk fluid is possible, but we observe no such nucleation in the simulations presented here.

\section{\label{sec:theory}Theory}
\subsection{\label{sec:gasliq}Free energy landscape}

Although this work is concerned with the dynamics of phase transitions, it is possible to gain a surprising level of insight into the expected behaviour from thermodynamic considerations alone \cite{Poon1999, Poon2000, Renth2001}, by examining the free energy landscape of the system. To that end, we now calculate approximate fluid and crystal free energy curves for our system, in the monodisperse case, by perturbation to a hard-sphere reference system.

The free energy density $f$ is given by:

\begin{equation}
f \approx  f_{\textrm{HS}} + \frac{\rho^2}{2} \int \! 4\pi r^{2} g_{\textrm{HS}}(r) U(r) \, dr
\label{eqn:hsperturbation}
\end{equation}

\noindent where $U(r) = -u$ and $\rho$ is the number density. The integral is over the square well range. For the hard-sphere contribution $f_{\textrm{HS}}$, we use the Carnahan-Starling free energy density \cite{Carnahan1969} for the fluid, and Hall's expression for the crystal \cite{Hall1972}. For the hard-sphere radial distribution function $g_{\textrm{HS}}(r)$, we use the Percus-Yevick expression in the fluid \cite{Stell1963} (using the analytical form for the first neighbour shell given in Ref.~\cite{Trokhymchuk2005}) and that of Choi et al. for the crystal \cite{Choi1991}.

\begin{figure}
\centering
\includegraphics[width=8.6cm]{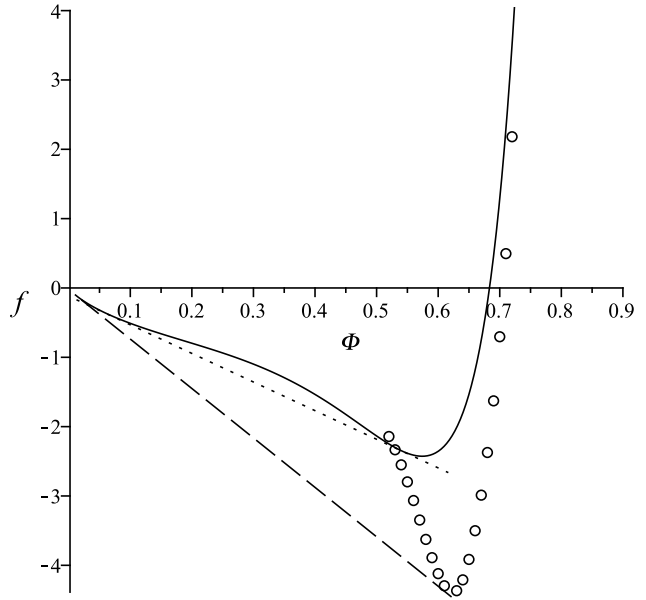}
\caption{\label{freeenergy}Free energy density $f$ as a function of volume fraction $\phi$ for a monodisperse system with $u = 1.82$ and $\lambda = 1.15$. The solid curve shows the fluid branch of the free energy while the circles show the crystal branch. The dotted line indicates the common tangent for metastable gas-liquid coexistence, and the dashed line shows the tangent for the equilibrium crystal-gas coexistence. There is no common tangent between the liquid and crystal phases.}
\end{figure}

FIG.~\ref{freeenergy} shows the resulting free energy densities plotted as functions of volume fraction for the $u = 1.82$ case (inside the G-L binodal, for our value of $\phi_{p}$, at point 1 in FIG.~\ref{LGK}). Allowed coexistences are given by the common tangent construction. Although quantitative agreement with the simulation data of Ref.~\cite{Liu2005} (in terms of the coexistence volume fractions of the gas and liquid phases) is relatively poor, some important qualitative features are present. The common tangent linking the minimum of the crystal branch with the gas is lower than that linking the liquid and gas -- the crystal-gas coexistence therefore has a lower overall free energy and is hence the equilibrium state, while the gas-liquid coexistence is metastable. Furthermore, there is \textit{no common tangent} between the crystal and liquid, which means that on the basis of free energy considerations, the crystal cannot locally coexist with the metastable liquid. 

The lack of a crystal-liquid common tangent implies a growth scenario like that proposed for the experimental colloid-polymer mixture observations in Refs.~\cite{EvansPoonRenth, Renth2001}, in which growing crystallites are coated by a layer of colloidal gas because of their inability to coexist with the metastable liquid. This `boiled-egg crystal' should deplete the surrounding bulk fluid until the required (equilibrium) crystal-gas coexistence is achieved overall. Our simulation setup allows us to model the effects of a metastable gas layer on the growth of a single crystal in a system for which we have a calculation of the corresponding free energy landscape, allowing new insight into some of the physics described in Refs.~\cite{EvansPoonRenth, Renth2001}. 

\subsection{\label{sec:split}Split interfaces}

The expected growth scenario of the `boiled-egg crystal' described in Section~\ref{sec:gasliq} is that of a `split interface'. That is, the crystal should form an interface with a gaseous region, which in turn forms an interface with the metastable liquid, since the crystal and metastable liquid cannot coexist on the free energy grounds discussed above. Such interfaces and a method for predicting their growth rates are described in Refs.~\cite{Evans1997, EvansPoonCates}, wherein the example used was of a crystal-liquid-gas interface as opposed to the crystal-gas-liquid interface we expect here. We now make use of that theory to predict the evolution of the split interface that should be formed when our system is inside the G-L binodal.

\begin{figure}
\centering
\includegraphics[width=8.6cm]{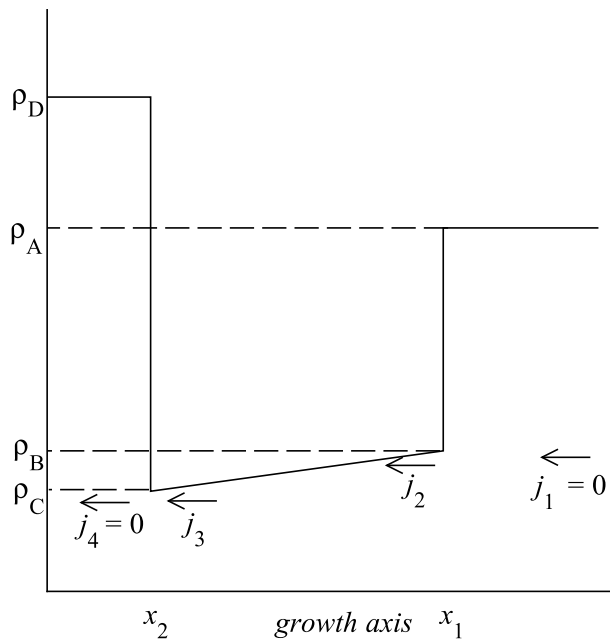}
\caption{\label{profile}Schematic density profile of the expected crystal-gas-liquid split interface inside the G-L binodal, where the crystal cannot coexist with the metastable liquid. The phases from left to right are the crystal, gas, and metastable liquid. The gradient in the gas region is drawn uniform, as is assumed in our calculations.}
\end{figure}

As shown in FIG.~\ref{profile}, we assume the growth scenario to be such that the phases on each side of the interfaces are at their `correct' densities according to the equilibrium phase diagram (FIG.~\ref{LGK}), the densities $\rho_{A}, \rho_{B}, \rho_{C}, \rho_{D}$ corresponding respectively to: the metastable liquid, the metastable gas, the stable gas, and the stable crystal. The assumption is therefore one of local equilibrium at the interfaces. 

Let $x_{1}(t)$ be the position of the gas-liquid interface, and $x_{2}(t)$ that of the crystal-gas interface. The respective fluxes on to and away from the interfaces are $j_{1}$, $j_{2}$, $j_3$ and $j_4$, as shown in FIG.~\ref{profile}. We assume $j_1 = 0$, i.e.\ that the liquid exists uniformly at its metastable density, and $j_4 = 0$ within the crystal region. The remaining fluxes can then be related to the interface speeds and the densities $\rho_{i}$:

\begin{align}
\centering
j_{2} &= \dot{x}_{1}(t) (\rho_{A} - \rho_{B}) \label{eqn:interfacespeed1} \\
j_{3} &= \dot{x}_{2}(t) (\rho_{D} - \rho_{C})
\label{eqn:interfacespeed2}
\end{align}

\noindent We now make the simplifying approximation that the gas region supports a uniform gradient between the densities $\rho_{B}$ and $\rho_{C}$, as is shown in FIG.~\ref{profile}, and further that the diffusion constant therein is equal to the ideal Stokes-Einstein diffusion constant $D_{0}$ due to the low density of the gas. Therefore: 

\begin{equation}
j_{2} \approx j_{3} \approx D_{0}\frac{\rho_{B} - \rho_{C}}{\Delta x }
\label{eqn:equalfluxes}
\end{equation}

\noindent where $\Delta x = x_{1}(t) - x_{2}(t)$. Substituting into Equations~\ref{eqn:interfacespeed1} and \ref{eqn:interfacespeed2} and separating variables, with the necessary constants of integration given by the initial conditions $x_{1}(t) = x_{2}(t) = 0$, we find expressions for the interface positions:

\begin{align}
x_{1}(t) = \sqrt{2D_{0}\frac{\beta^{2}}{\gamma}t}  + \sqrt{2D_{0}\gamma t}   \label{eqn:interfacepos1} \\
x_{2}(t) = \sqrt{2D_{0}\frac{\beta^{2}}{\gamma}t} 
\label{eqn:interfacepos2}
\end{align}

\noindent where we have defined $\beta \equiv (\rho_{B}-\rho_{C})/(\rho_{D}-\rho_{C})$ and $\gamma \equiv  (\rho_{B}-\rho_{C})/(\rho_{A}-\rho_{B}) -  (\rho_{B}-\rho_{C})/(\rho_{D}-\rho_{C})$. 

\begin{figure}
\centering
\includegraphics[width=8.6cm]{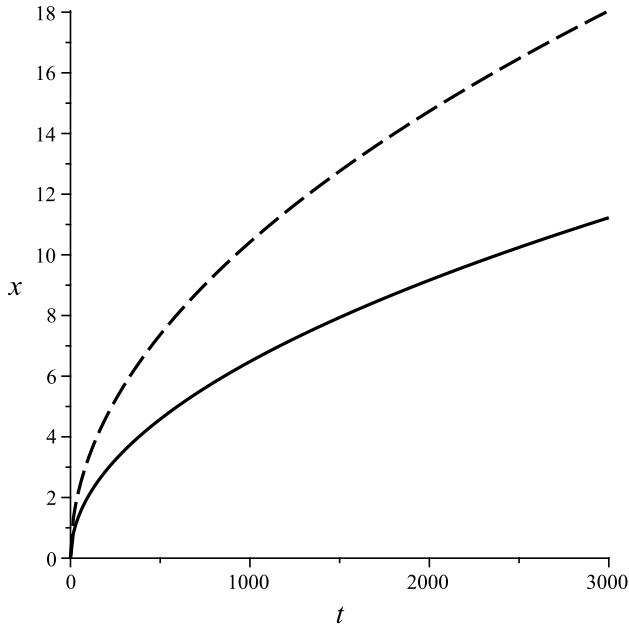}
\caption{\label{growthcurves}The predicted evolution of the crystal-gas (solid line) and gas-liquid (dashed line) interfaces for a split crystal-gas-liquid interface when our system is inside the G-L binodal ($u = 1.82$, $\lambda = 1.15$).}
\end{figure}

To make contact with our simulations, we read off the values of $\rho_{i}$ from FIG.~\ref{LGK} at the appropriate effective temperature $T_{\textrm{eff}} = 1 / 1.82 \approx 0.55$ for point 1, inside the G-L binodal, and use the Stokes-Einstein diffusion constant defined by our choice of time unit, $D_{0} = 1/6$. The resulting interface positions through time are shown in FIG.~\ref{growthcurves}, and in the next section are compared with simulation data. The characteristic $t^{1/2}$ diffusive growth is apparent, with the gas-liquid interface leading the crystal-gas interface as expected. 

We note at this stage that the crystal-gas-liquid split interface scenario is closely analogous to the crystal-liquid-gas interfaces described in Refs.~\cite{Evans1997, EvansPoonCates}. The only difference is that in the present work, the `pivot' phase which coexists locally with both others (and therefore coats the crystal) is the gas, i.e.\ \textit{is the equilibrium phase}, whereas in the aforementioned work it is the metastable, i.e.\  nonequilibrium liquid phase.

\section{\label{sec:results}Results and Discussion}

In this section, we examine the simulation results in terms of their time-dependent concentration profiles and in terms of their crystal growth rates, elucidating and discussing the separate and combined effects of metastability and polydispersity on the growth process. Then, in Section~\ref{sec:fractionation}, we outline and briefly test a possible explanation for our findings in terms of fractionation and rearrangement at the crystal interface.

We present results for polydispersities of $\sigma = 0$ (monodisperse), $\sigma = 0.03$ and $\sigma = 0.06$. The $\phi$ and $u$ coordinates used are marked as points `1' and `2' on the monodisperse phase diagram, FIG.~\ref{LGK}. We use square well depths of $u=1.82$ (point 1, inside the G-L binodal, referred to as `gas-mediated') or $u=1.54$ (point 2, outside the G-L binodal, referred to as `gas-free'). The combinations of $u$ and $\sigma$ give 6 state points in total.  At each state point, 6 independent initial configurations were used. 

\subsection{\label{sec:concentration}Concentration profiles -- monodisperse}

The clearest way to visualise crystal growth over the whole simulation time is by plotting the time-dependent concentration profile. To achieve this, we plot the `area fraction' $\phi _{\textrm{area}}$ of particles intersecting planes at a given $x$ coordinate, corresponding to a local volume fraction in an infinitely narrow interval $\delta x$. Since the simulation geometry is such that the crystal interface advances along the $x$ axis, these plots are a powerful way of observing the time evolution of concentration differences along the direction of growth while averaging over the axes perpendicular to it. 

In this section, we qualitatively compare representative concentration profiles in the monodisperse and $\sigma = 0.06$ cases. Except where noted, the phenomena described occurred similarly in all trajectories of the state point in question.

\begin{figure}
\centering
\includegraphics[width=8.6cm]{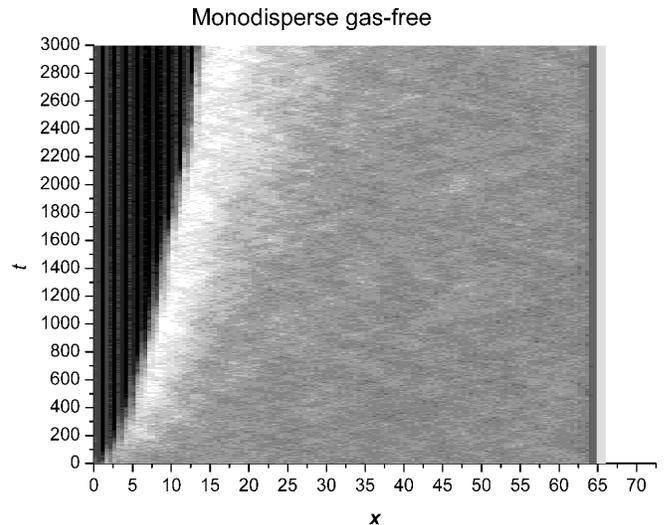}
\caption{\label{mono7a}Time-dependent concentration profile along the $x$ axis for one of the monodisperse gas-free simulations. The greyscale indicates the local volume fraction, ranging from $\phi < 0.06$ (white) to $\phi > 0.676$ (black). As the simulation progresses, the crystal advances along the $x$ axis, depleting the fluid in front of the interface.}
\end{figure}

\begin{figure}
\centering
\includegraphics[width=8.6cm]{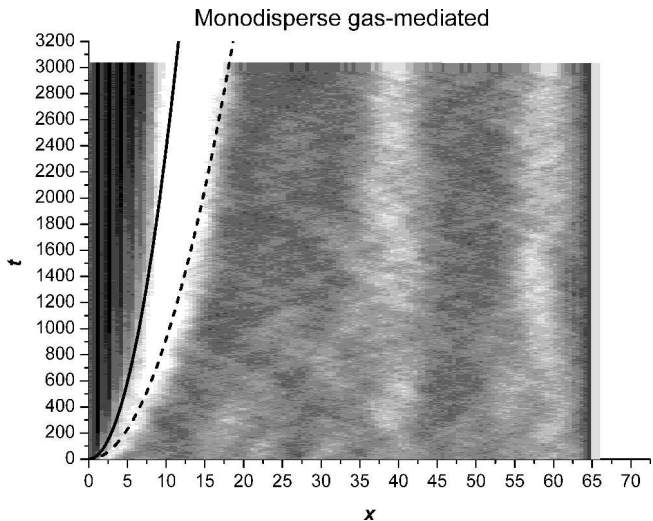}
\caption{\label{mono6a}As FIG.~\ref{mono7a}, for a monodisperse gas-mediated simulation. Gas-liquid separation takes place in the bulk, and the advancing crystal is coated by a distinct gas layer which advances ahead of it. The theoretical crystal-gas and gas-liquid interface curves from FIG.~\ref{growthcurves} are rotated and superimposed, showing approximate agreement with the data. }
\end{figure}

Let us consider the monodisperse case first. FIGs.~\ref{mono7a} and \ref{mono6a} show example trajectories for simulations at the gas-free (point 2 on FIG.~\ref{LGK}) and gas-mediated (point 1 on FIG.~\ref{LGK}) state points respectively. In the gas-free case, the growth scenario is, as expected, relatively simple. The bulk fluid remains homogeneous, since we are above the critical $T_{\textrm{eff}}$ for G-L separation. The crystal template causes the growth of a single crystalline region with an average volume fraction of $\phi \approx 0.6$, which is the expected equilibrium volume fraction for the crystal \cite{Liu2005}. The crystal, being at a higher volume fraction than the bulk fluid, depletes its surroundings of particles, resulting in a concentration gradient between the local fluid and the bulk; it is this gradient which transports particles toward the crystal so that it can continue to grow. 

\begin{figure}
\centering
\includegraphics[width=8.6cm]{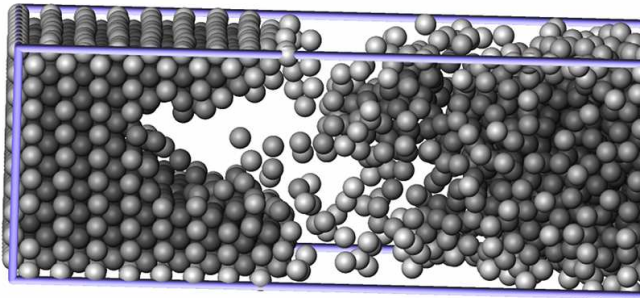}
\caption{\label{dendritic}Snapshot of the crystal interface for one particular monodisperse gas-mediated  (i.e.\  inside the G-L binodal) trajectory in which the crystal happened to exhibit dendritic growth around an impinging gas bubble. Such dendritic growth was not generally observed.}
\end{figure}

Inside the G-L binodal (FIG.~\ref{mono6a}), the scenario is quite different. Firstly, there is clearly some G-L separation taking place in the bulk. Focusing next on the crystal template at $x=0$, we can see that a region of low density gas forms in front of the crystal almost immediately, shielding it from the liquid with which it cannot locally coexist according to the free energy considerations in Section~\ref{sec:gasliq}. This is in contrast to FIG.~\ref{mono7a}, in which the fluid immediately next to the crystal template retains its density for quite some time until the growing crystal depletes it of particles. Note also that whereas the depleted region in FIG.~\ref{mono7a} fades smoothly into the higher density bulk fluid, the gas next to the crystal in FIG.~\ref{mono6a} forms a substantially sharper interface with the fluid next to it, indicating a distinct phase boundary between it and the metastable liquid. The formation of this well-defined, macroscopic gas layer shielding the crystal is consistent with the free energy considerations above, and with the experimental observations of \cite{Renth2001}. 

The speed of advance of the crystal-gas and gas-liquid interfaces is in approximate agreement (to within 1 particle diameter) with the theoretical prediction in Section~\ref{sec:split} (FIG.~\ref{growthcurves}), which we have rotated and superimposed on FIG.~\ref{mono6a} for comparison. This agreement is particularly satisfying given that the input densities $\rho_{i}$ required for the theoretical calculation in Section~\ref{sec:split} were taken from the equilibrium phase diagram, so that there are no free fitting parameters. The accuracy of the prediction thus gives credence to the assumptions of local equilibration of the interfaces and of essentially ideal gas-like Stokes-Einstein diffusion in the gas region. We note also that the theoretical prediction does not take account of the `start-up' stage when the gas layer is just forming, since it assumes a gas layer (albeit one of zero size) to be present from $t = 0$ -- this may explain the slight discrepancy at early times.

Finally, it is interesting to note that for some trajectories, the crystal can grow incomplete layers, as shown in FIG.~\ref{dendritic}. This is the result of dendritic growth of the crystal around an impinging gas bubble. Such growth was not found to be generally present in the other trajectories at this state point -- the crystal typically formed complete layers covered by a layer of gas spanning the whole width of the simulation cell. 

\subsection{\label{sec:concentrationpoly}Concentration profiles -- polydisperse}

\begin{figure}[ht]
\centering
\includegraphics[width=8.6cm]{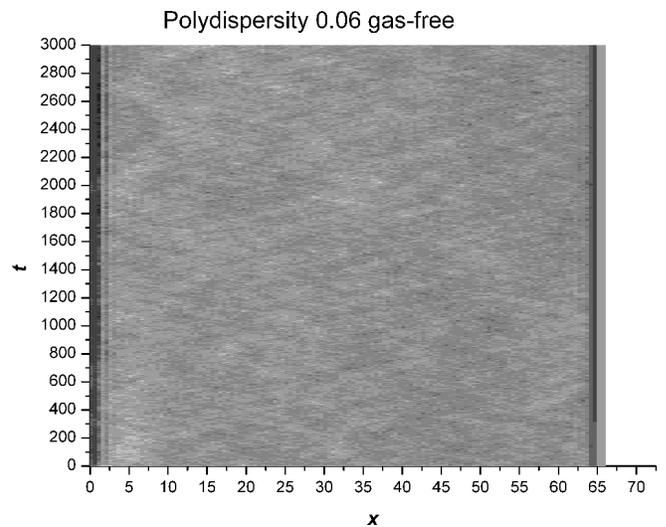}
\caption{\label{poly7a}As FIG.~\ref{mono7a}, for a gas-free system of polydispersity $\sigma = 0.06$. As in the monodisperse case, the bulk fluid remains homogeneous. However, the crystal template does not induce any growth on the timescale simulated.}
\end{figure}

\begin{figure}[ht]
\centering
\includegraphics[width=8.6cm]{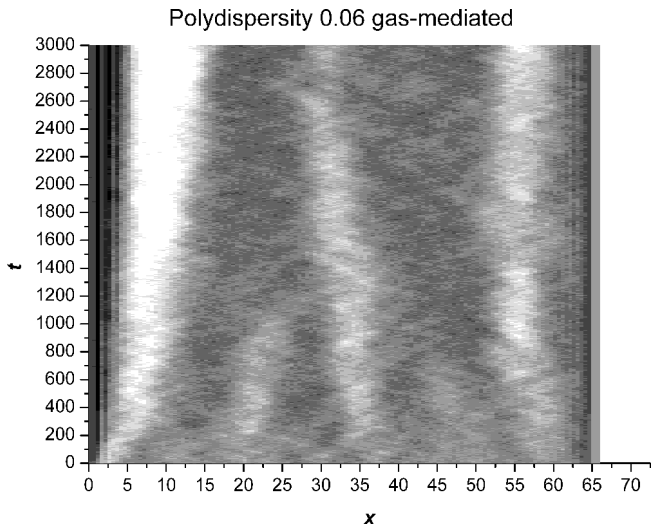}
\caption{\label{poly6a}As FIG.~\ref{mono7a}, for a gas-mediated system of polydispersity $\sigma = 0.06$. The bulk fluid shows G-L separation, and the gas layer is present in front of the crystal. In contrast to FIG.~\ref{poly7a}, the crystal is able to grow, although more slowly than in the corresponding monodisperse case (FIG.~\ref{mono6a}).}
\end{figure}

We now discuss the polydisperse $\sigma = 0.06$ case. FIG.~\ref{poly7a} shows the gas-free scenario, corresponding again to point 2 on FIG.~\ref{LGK}. On the simulated timescale, essentially no crystal growth is seen to take place, except perhaps for a slight `wetting' of the template by ordered particles. As in the monodisperse case, the bulk fluid remains homogeneous, indicating that the polydispersity has not altered the location of the critical $T_{\textrm{eff}}$ in such a way as might bring this state point inside the G-L binodal. 

\begin{figure*}[ht]
\centering
\includegraphics[width=17.6cm]{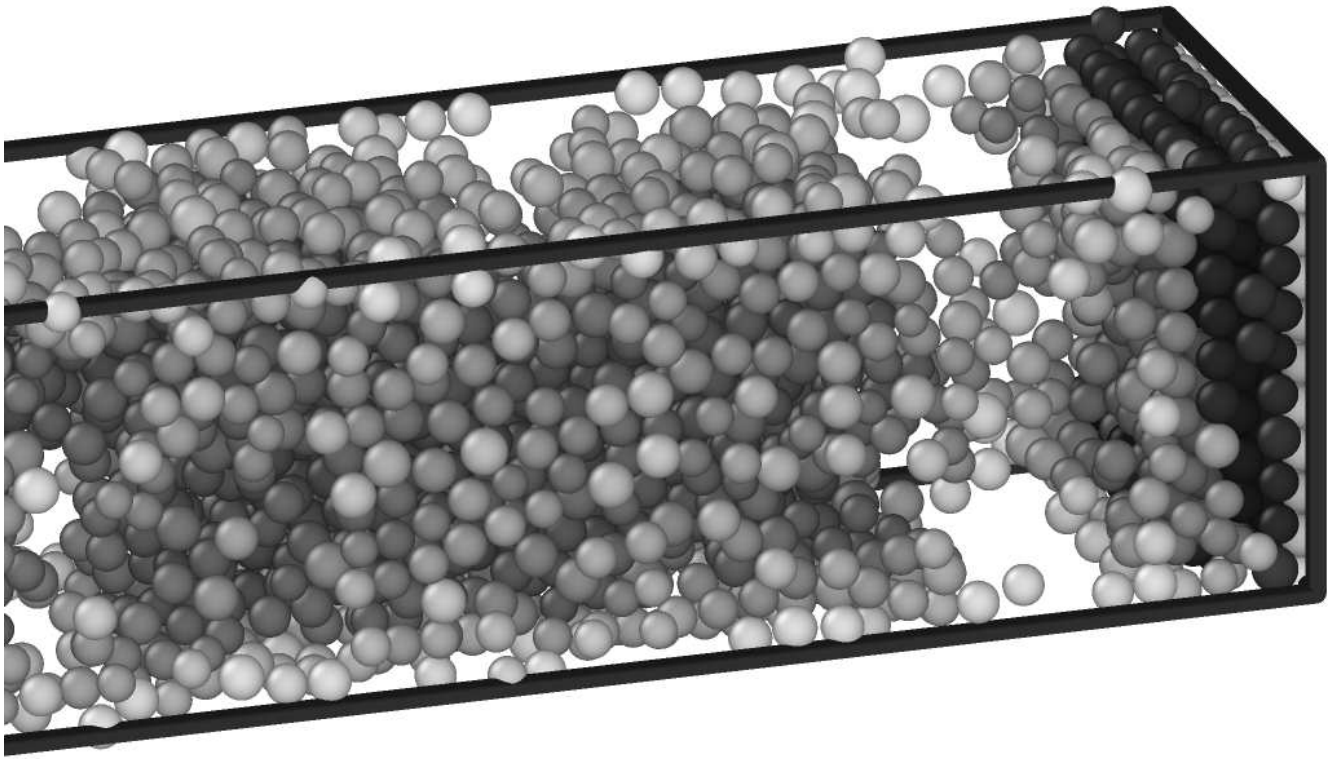}
\caption{\label{flagging}Illustrative snapshot of a polydisperse $\sigma = 0.06$ gas-mediated trajectory at $t \approx 900$, showing the gas-coated crystal region. Particles flagged crystalline by the bond order parameter algorithm described in Section~\ref{sec:geometry} are shown in black, all others are light grey.}
\end{figure*}

Moving to the gas-mediated case inside the binodal (at point 1 on FIG.~\ref{LGK}), for which a trajectory is shown in FIG.~\ref{poly6a}, we can see that the crystal is able to grow substantially despite the polydispersity. As in the corresponding monodisperse case (FIG.~\ref{mono6a}), the bulk fluid separates into gas and liquid regions, and a gas layer covers the crystal as it grows. An illustrative snapshot of the gas-coated polydisperse crystal is shown in FIG.~\ref{flagging}.

To summarise the qualitative observations in this section: the crystal template successfully induces a physically realistic growth process, in which the crystal depletes its surroundings and forms ordered, high density layers. When the system is inside the G-L binodal (at point 1 in FIG.~\ref{LGK}), a gas layer forms in front of the crystal, shielding the crystal from the metastable liquid. This is exactly the scenario predicted for our system in Section~\ref{sec:gasliq} and proposed in Refs.~\cite{EvansPoonRenth, Renth2001}. The presence of polydispersity $\sigma = 0.06$ substantially slowed the crystal growth overall. In the gas-free case (point 2 on FIG.~\ref{LGK}) the crystal showed barely any growth on the simulated timescale, whereas the gas-mediated case (point 1) showed substantial crystal growth despite the polydispersity.

\subsection{\label{sec:growthrate}Monodisperse crystal growth rates -- the effect of metastability}

\begin{figure}
\centering
\includegraphics[width=8.6cm]{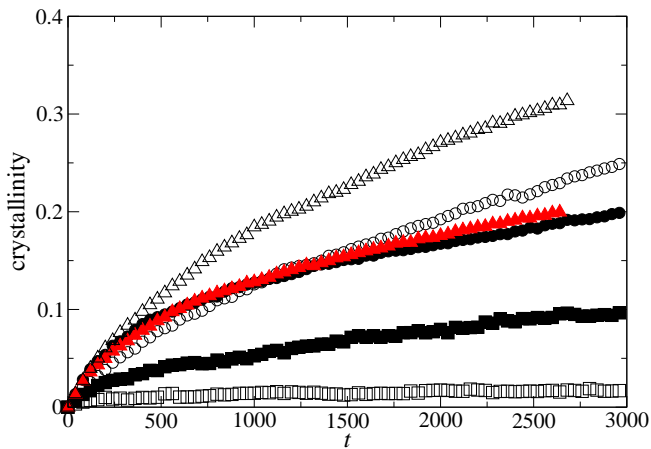}
\caption{\label{growthrates}(colour online) Crystallinity through time for the 6 state points studied, in each case averaged over 6 independent realisations. Filled symbols: gas-mediated (point 1 on FIG.~\ref{LGK});  open symbols: gas-free (point 2 on FIG.~\ref{LGK}). Triangles: monodisperse; circles: $\sigma = 0.03$; squares: $\sigma = 0.06$. The standard errors are approximately the size of the symbols. The filled triangles (monodisperse gas-mediated) are shown in grey (red online) to distinguish them from the filled circles ($\sigma = 0.03$ gas-mediated).}
\end{figure}

Having summarised the qualitative features of the monodisperse and $\sigma = 0.06$ simulations, we next quantitatively examine the effect of those features on the crystal growth rate. This is done by plotting the crystallinity (the proportion of particles flagged as crystalline, according to the bond order criterion described in Section~\ref{sec:geometry}) through time. FIG.~\ref{growthrates} shows the crystal growth at each of the 6 state points studied, each averaged over the 6 independent realisations of the state point.

In the monodisperse case, the effect of metastability alone is apparent. The gas layer in front of the crystal, which forms when we are at point 1 in FIG.~\ref{LGK}, significantly slows crystal growth compared to the gas-free case, corresponding to point 2 in FIG.~\ref{LGK}. 

Since the crystal exists at a higher density than the rest of the system, as soon as the immediate surroundings are depleted of particles, particles must be transported toward the crystal from the bulk in order for it to grow. This transport takes place via collective diffusion down a concentration gradient between the low concentration near the crystal and the relatively higher concentration away from the crystal. The diffusion is described by Fick's law,

\begin{equation}
J = - D\nabla \phi
\label{eqn:fick}
\end{equation}

\noindent in which $J$ is the concentration flux, $\phi$ is the local concentration and $D$ is the (concentration-dependent) collective diffusion coefficient. Note that the collective diffusion coefficient, in contrast to the self diffusion coefficient, increases with $\phi$ \cite{Tirado2003}. 

We now consider the phase diagram for our system, as shown in FIG.~\ref{LGK} (c). In the absence of the gas layer, and assuming that the interface is locally equilibrated, a concentration gradient exists between $\phi \approx 0.05$ at the interface and $\phi_p = 0.34$ in the bulk. However, inside the G-L binodal, when the crystal is covered by a distinct gas layer, the relevant concentration gradient is that which exists \textit{across the gas layer itself} -- the gas-liquid interface beyond truncates the concentration gradient. The gradient in the gas layer is between the equilibrium gas ($\phi \approx 0.01$) and the slightly higher concentration metastable gas at $\phi \approx 0.05$. The gas layer therefore results in a significantly smaller concentration difference being spread across a comparable distance -- from FIGs.~\ref{mono7a} and \ref{mono6a}, the gas layer is similar in size to the depletion region in the gas-free case.

Therefore, the low-concentration gas layer affects particle transport in two ways, relative to the gas-free case: (a) the collective diffusion coefficient $D$ is lower, due to reduced concentration; (b) the gas's low concentration and macroscopic size mean it can only support a relatively weak concentration gradient in comparison to the gas-free case. As evidenced in the monodisperse crystal growth rates (FIG.~\ref{growthrates}), the net result is that the flux onto the crystal, and therefore the interface speed, is lower when the gas is present. Although, on thermodynamic grounds, the driving force for crystallisation appears higher inside the G-L binodal, the growth of a given crystal is slowed due to the gas layer's effects on particle transport to the interface.

The effect of the gas layer in the monodisperse case is comparable to that of the liquid layer in Refs.~\cite{Evans1997, EvansPoonCates}, which was also found to inhibit crystal growth. There, as here, the crystal is coated by a phase which advances ahead of the crystal and slows down its growth. It is interesting that this comparison holds, given that in our case the crystal is being coated by the \textit{equilibrium} gas, whereas Refs.~\cite{Evans1997, EvansPoonCates} concern a crystal coated by a \textit{nonequilibrium} liquid.

\subsection{\label{sec:growthratepoly}Polydisperse crystal growth rates}

We consider first the highest polydispersity studied here, $\sigma = 0.06$. As shown in FIG.~\ref{growthrates} and in agreement with the observations in Section~\ref{sec:concentrationpoly}, the gas-free state point now shows only very slight growth on the timescale simulated, whereas the gas-mediated state point shows significant crystal growth (although less than in either of the monodisperse systems). This is a \textit{qualitative difference} compared to the monodisperse case, in which the gas layer instead strongly slows the crystal growth. That is, the presence of polydispersity qualitatively alters the effect of the metastable gas layer on crystal growth. 

In the $\sigma = 0.03$ case, the gas-mediated system is initially slightly faster, before being overtaken by the gas-free system by around $t = 1000$. This is suggestive of some kind of crossover phenomenon between whatever factors are dominant in the fully monodisperse and $\sigma = 0.06$ cases.

Given the relative proximity of state point 1 to the gas-liquid critical point (FIG.~\ref{LGK}), the possible influence of critical fluctuations was considered. We therefore tested a stronger well parameter, $u=1.9$, $T_{\textrm{eff}} = 0.526$, at $\sigma = 0.06$, to see any effect of moving further away from the gas-liquid critical point. The crystal growth rate was unaffected within error, suggesting that the critical point is not close enough to state point 1 for critical fluctuations to have an effect. In addition, the observed gas-liquid interface on the far side of the gas layer appears sharp -- we do not observe the large fluctuations characteristic of critical phenomena.

Also for $\sigma = 0.06$, moving the (gas-free) state point 2 to $T_\textrm{eff} = 0.6$ (to increase the driving force for crystallisation) slightly speeded up growth, but it was still much slower than in the gas-mediated case. Further study in this direction is left to future work.

\section{\label{sec:fractionation}Crystallisation mechanism}

\subsection{\label{sec:proposal}Fractionation}

The results presented so far show that metastable gas-liquid separation can affect the rate of crystal growth, because it results in a gaseous layer coating the crystal as it grows. However, the qualitative nature of that effect depends strongly upon polydispersity. In the monodisperse case, the resultant gas layer impedes crystal growth by reducing the efficiency of particle transport to the crystal. At $\sigma = 0.06$, the metastable separation instead strongly enhances crystal growth. It is this effect of polydispersity on the crystal growth mechanism that we now discuss.

We propose an explanation in terms of a local fractionation process at the interface of the polydisperse crystal. In size polydisperse systems, it is well known that phase separation is typically associated with some degree of thermodynamically-driven fractionation in which one phase e.g.\ contains on average larger or smaller particles than another, changes its overall polydispersity, etc. Such fractionation has been predicted theoretically \cite{Evans2000, Fasolo2004}, and observed experimentally \cite{Fairhurst2004} and in simulation \cite{Williamson2012}. Of course, this kind of fractionation pertains to the \textit{equilibrium} phase composition and may not be fully achieved in real systems on accessible timescales due to kinetic factors. This is a particularly important consideration for the crystal phase, where particles are essentially stationary once incorporated and, thereafter, the long-range particle transport required for fractionation is facilitated only by the presence of defects. In general, phase composition is expected to relax slowly in comparison with overall density so that fractionation may lag behind phase separation somewhat \cite{Warren1999}, leading in the crystal phase to the `freezing-in' of a nonequilibrium composition \cite{Evans2001}. (Note, however, that recent work by us has shown that \textit{some} fractionation is possible in the very early stages of gas-liquid separation \cite{Williamson2012}.)

Nevertheless, we would expect that fractionation \textit{at the interface} of the growing crystal, where particles are still mobile and may be easily exchanged with the fluid, is quite feasible. Indeed, the slowing effect of polydispersity on crystallisation is generally taken as evidence that fractionation is involved to some extent \cite{Martin2003, Schope2007}. We propose that such interfacial fractionation takes place not just in nucleation but during the subsequent growth of the crystal, so that it is facilitated by the low-density gas layer that forms in front of the crystal when inside the G-L binodal. In contrast to the collective diffusion discussed in Section~\ref{sec:growthrate}, the self diffusion required for fractionation is hindered by particle density, so that fractionation would be frustrated outside the G-L binodal, where the fluid side of the interface is at a relatively high density compared to that in the gas-mediated case.  

In addition to an \textit{overall} preference for a narrower size distribution, it seems reasonable to suppose that the crystal may be selective in some way as to which particles are incorporated where, on a local basis. For instance, it may be frustrated by the presence of regions occupied solely by very large or small particles. Whatever the nature of any local size ordering, we expect that the crystal should have some kind of preference as to how particle size is distributed within it, given that the largest and smallest particles in the $\sigma = 0.06$ system differ in size by around $40\%$ of the mean. This local ordering is conceived of as being in addition to any fractionation in terms of the overall balance of particle sizes between the phases.

From the outset, we stress that a full explanation of the results in the previous section requires further work to elucidate and test other possible contributing factors, such as the influence of polydispersity on the equilibrium phase diagram. The proposals here are motivated by the dynamical nature of our simulation approach and by the intuition that the increased diffusivity of particles in the gaseous layer would help any dynamical sorting processes at the crystal interface, a consideration which we expect to be a significant aspect of any full explanation of the results. In the following, we perform further analysis on the simulation data in an attempt to detect the presence of fractionation processes.

\subsection{\label{sec:poly}Polydispersity of the crystal}

The results we now present concern the gas-mediated (point 1 in FIG.~\ref{LGK}) and gas-free (point 2 in FIG.~\ref{LGK}) $\sigma = 0.06$ simulations. To ensure good sampling in the crystal, we allowed these to run up to $t \approx 9000$. By this time, the crystal interface in the gas-mediated simulations reached $x \approx 8$, while the gas-free simulations had only grown to $x \approx 3.5$. We check for fractionation by measuring the mean diameter $\langle d \rangle _\textrm{cryst}$ and polydispersity $\sigma_\textrm{cryst}$ in the crystal. For the purposes of the present analysis, we identify the crystal interface at a given time as the centre of an averaging window of length $1.5$ in $x$ in which, moving through the system from $x = 0$, the average local volume fraction first drops below the parent value $\phi = 0.34$, due to the depleted region in front of the crystal. The identified interface positions are checked against the concentration profiles (e.g.\ FIG.~\ref{mono6a}) and against direct visual observations, and the region lower in $x$ than the interface is considered to be the crystalline region. For the case of FIG.~\ref{dendritic}, in which the crystal interface is not flat, our definition means that only full or nearly-full crystal layers fall into the `crystalline region'.

\begin{figure}
\centering
\includegraphics[width=8.6cm]{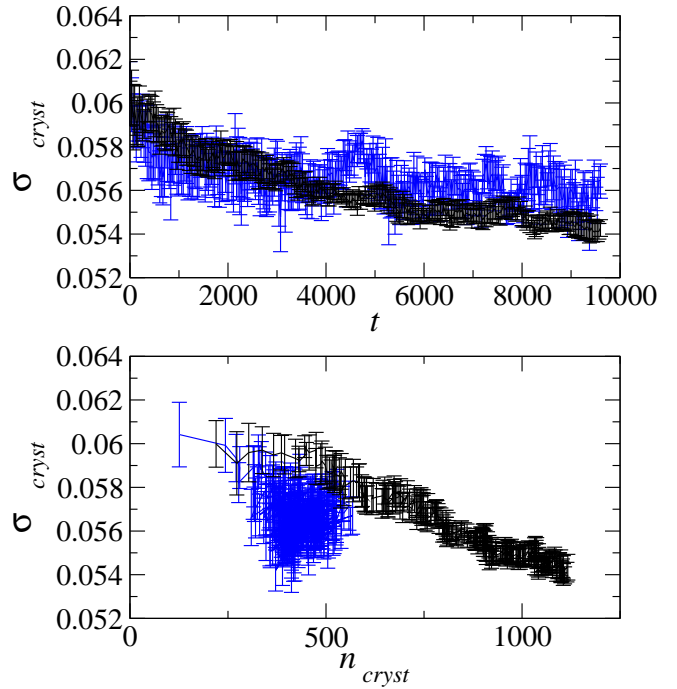}
\caption{\label{polyreduce}(colour online) Crystal polydispersity $\sigma_\textrm{cryst}$ as a function of $t$ (upper pane) and $n_\textrm{cryst}$ (lower pane) for the gas-mediated (black) and gas-free (grey, blue online, larger error bars) crystals. The error on the $n_\textrm{cryst}$ axis is approximately 30.}
\end{figure}

To within statistical error, $\langle d \rangle _\textrm{cryst}$ was equal to the parental mean, $\langle d \rangle$. However, changes in $\sigma_\textrm{cryst}$ were detected. In FIG.~\ref{polyreduce}, the evolution of $\sigma_\textrm{cryst}$ in the gas-mediated and gas-free crystals is shown, as a function of time and of $n_\textrm{cryst}$, the mean number of particles in the crystal. In the gas-mediated case, there is a clear reduction in the polydispersity of the crystal compared to the parental $\sigma = 0.06$, demonstrating that the crystal is selecting a narrower subset of the parent size distribution, eventually attaining a value of $\sigma_\textrm{cryst} \approx 0.054$ at the end of the simulated time. This is qualitatively consistent with Fasolo and Sollich's equilibrium calculations on hard-spheres, in which even a small parental polydispersity was found to lead to reduced polydispersity in a crystal coexisting with a fluid \cite{Fasolo2004}. This measurement in itself is significant: to our knowledge, such fractionation has not previously been measured in the dynamical crystal growth of polydisperse spherical colloids (an analogous example for colloidal platelets exists in Ref.~\cite{Byelov2010}), although it is proposed as an explanation of slow crystallisation in such systems \cite{Martin2003, Schope2007}. 

The data for the gas-free crystal are subject to significantly larger error, because very little crystal has actually formed, but appear to be consistent with a small reduction in polydispersity corresponding to the frustrated crystal growth. In any case, it is clear that the successful growth of a crystal involves, as expected from equilibrium work \cite{Fasolo2004, Sollich2011}, a measurable reduction in polydispersity from the bulk fluid, indicating a fractionation process which would be enhanced by the gas layer.

\begin{figure}[ht]
\centering
\includegraphics[width=8.6cm]{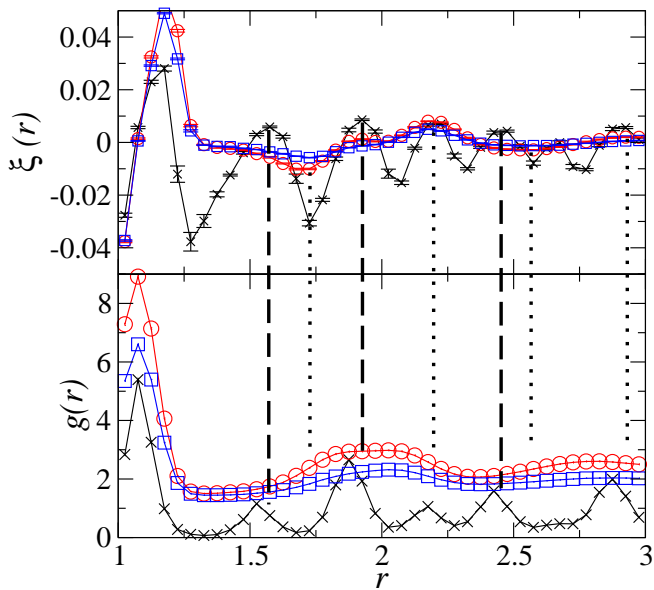}
\caption{\label{localcorr}(colour online) Upper pane:\ the local size correlation function $\xi(r)$ as defined in the text, measured at late times for systems of polydispersity $\sigma = 0.06$. Black crosses indicate the gas-mediated crystal, red circles the gas-mediated outside-crystal region (comprising gas and liquid coexisting), and blue squares the gas-free outside-crystal region (a homogeneous fluid). Lower pane:\ the corresponding radial distribution functions measured in the same systems. Colours and symbols as for the upper pane. Both the outside-crystal $g(r)$ datasets have been multiplied by a factor of 2 for ease of visualisation. Lines between the data points are guides to the eye. Vertical lines linking the panes are provided to show the relationship between the functions -- dashed lines for the crystal, and dotted lines for the two fluid datasets, whose dependence on $r$ is essentially identical. The data points are placed in the centre of each bin in the $r$ axis.}
\end{figure}

\subsection{\label{sec:localcorr}Local size correlations}

To check for local size ordering, in addition to the overall preference for a narrower particle distribution, we define a local diameter-diameter correlation function $\xi(r)$:

\begin{equation}
\xi(r) = \langle d_{i} d_{j}\rangle_{\textit{r}} - \langle d_{i} \rangle^{2}
\label{eqn:corrfunc}
\end{equation}

\noindent where the averaging in the first term is over neighbours separated by a distance $r$. The function $\xi(r)$ is therefore analogous to a `radial distribution function,' but for size correlations, as opposed to density correlations, and tends to zero in the limit of an ideal system (i.e.\  one in which particles do not interact and `size' just becomes a label which does not affect the behaviour of the particles). 

Measurements are taken from the simulations by binning in $r$, and can be made within either the crystalline or amorphous region, to observe any differences induced by crystallisation. Then, the averaging over the second term in Equation~\ref{eqn:corrfunc} is over all particles \textit{in that phase}. We compare $\xi(r)$ in the gas-mediated $\sigma = 0.06$ crystal with that outside the crystal in the same system (remember that in the gas-mediated case, `outside the crystal' refers to a metastable coexistence of gas and liquid in the bulk). We also measure $\xi(r)$ outside the crystal in the gas-free case, i.e.\  a homogeneous fluid of $\sigma = 0.06$. 

The results of this analysis are shown in FIG.~\ref{localcorr}, in which are also shown the measured radial distribution functions $g(r)$ corresponding to the $\xi(r)$ data. The data have been averaged within each simulation over $t = 8,000 - 9,000$, during which time both $\xi(r)$ and $g(r)$ appeared essentially static due to the very slow growth of the crystal (whose size scales as $t^{1/2}$), and then averaged over the independent simulations. 

The comparison reveals a fascinating `$g(r)$-like' appearance to the local size correlations both inside and outside the crystal. However, closer inspection reveals that whereas the oscillations in $\xi(r)$ for the crystal appear approximately in phase with those in its $g(r)$, those in the fluid -- both the homogeneous and G-L separating fluid cases -- are approximately in phase quadrature with $g(r)$.

Our proposed explanation for these findings is as follows. We will consider first the gas-mediated crystal, then the two fluid datasets. Before starting, we note that, in a polydisperse system, larger particles will tend to contribute a stronger signal to structural correlation functions. See, for instance, the measurements of partial static structure factors in Ref.~\cite{Weysser2010}. One may therefore be guided by this in interpreting $\xi(r)$:\ e.g.\ a positive value indicates a correlation between either `big-big' or `small-small' particle pairs, but will contain a stronger contribution from `big-big' pairs due to the stronger overall structuring of large particles. The terms `big' and `small' are here shorthand for particles greater or lesser in size than the mean $\langle d\rangle \equiv 1$.

\subsubsection{Crystal.~~}

In the crystal, the features in $\xi(r)$ more or less mimic those of $g(r)$, and the two are approximately in phase, aside from a slight offset for the first peak which decreases through the subsequent peaks. Where there is a peak in $g(r)$, i.e.\  a pair of neighbours is likely to be found separated by this distance, those particles on average show a slight positive correlation in terms of their size, indicating that the crystal structure has a preference for particles in the same region to be of similar size, presumably to minimise distortion of the lattice. Conversely, particle pairs separated by the unusual distances corresponding to minima in $g(r)$ have a rather strong negative correlation. This is consistent with a scenario in which these neighbour separations are associated with the presence of `wrong', i.e.\ unusually-sized, particles, for a given region of the crystal. They may, for instance, result from an interstitial defect, with a particularly small particle squeezing itself in amongst a region of average or slightly large on-lattice particles. 

The relationship between $\xi(r)$ and $g(r)$ in the crystal is therefore due to the need to minimise lattice distortions over small regions of the crystal, and the fact that unusual neighbour separations will tend to be associated with particles of unusual size for a given region distorting the crystal structure. We stress that this effect is \textit{in addition} to the overall preference for a reduced crystal polydispersity. Conceivably, the crystal could have reduced its polydispersity precisely in order to \textit{avoid} having to enact the local ordering observed. In fact, for the dynamically grown crystal here, a little of both effects seems to be required:\ the crystal reduces its polydispersity, but still `cares' about the local distribution of particle size in its lattice. Since the crystals grown in our dynamical simulations are not necessarily at equilibrium, it would be interesting to compare these results to those from an equilibrated polydisperse crystal. We intend to pursue this question in future work.

\subsubsection{Fluid.~~}

Both the homogeneous (gas-free, point 2 in FIG.~\ref{LGK}) and G-L separating (gas-mediated, point 1 in FIG.~\ref{LGK}) fluid regions show similar behaviour with respect to the $r$-dependence of $\xi(r)$ and $g(r)$, so we will not distinguish between them for the purposes of this analysis. In the fluid, $\xi(r)$ appears to be approximately in phase quadrature with $g(r)$, so that minima or maxima in $\xi(r)$ appear, respectively, halfway up or halfway down the slope around a peak in $g(r)$. 

We interpret this in terms of shells around a nominal test particle $i$, in analogy to the explanation of similar oscillations in terms of shells near a hard wall in Refs.~\cite{Pagonabarraga2000, Buzzacchi2004}. Consider that a single shell of surrounding particles corresponding to a peak in $g(r)$ may be roughly divided into those which are nearer than average and those which are further away than average. In order to most efficiently fill space, particles which are closer should be small, and those which are further away should be large. Put another way, particles which are further away should be further away \textit{because they are big}; placing small particles further away than is required by their size would waste space, as it were, reducing the packing efficiency and increasing the free energy of the fluid. Given the stronger structural signal from larger particles \cite{Weysser2010}, the data are dominated by the case where particle $i$ is big. Therefore, the smaller, closer, particles in the shell will on average contribute negatively to $\xi(r)$, resulting in a minimum therein at a value of $r$ slightly less than the peak in $g(r)$. Conversely, the larger, further away particles contribute positively, giving a peak in $\xi(r)$ just outside the peak in $g(r)$.

In the DFT study of Ref.~\cite{Pagonabarraga2000}, an observed phase quadrature between the size distribution near a wall and the mean density profile (roughly corresponding to our $\xi(r)$ and $g(r)$ functions) was explained in these terms. Our findings are therefore significant in showing that similar `local size segregation' appears to be present in the bulk fluid, rather than being solely the result of spatial inhomogeneity. We note for completeness that measuring $\xi(r)$ in a cubic system with no crystal template (so that our system is truly homogeneous and isotropic) had no significant effect on the behaviour of $\xi(r)$; our measurements do seem to capture the behaviour of a bulk fluid.  

Thus, the observed relationship between $\xi(r)$ and $g(r)$ in the fluid is a manifestation of the need for efficient packing in the fluid, coupled with the tendency for larger particles to contribute more strongly to the structuring signal. The fact that local size segregation previously observed near a hard wall appears to be manifest also in a bulk fluid is very interesting, and would seem to merit further investigation. In particular, we aim to investigate in future work whether this same effect is present in an equilibrium polydisperse fluid. 

\subsection{\label{sec:eq}Equilibrium insights}

Although our simulations are inherently not at equilibrium, it is important to consider the effects of polydispersity on the equilibrium phase diagram in interpreting the results. The most detailed investigations into polydisperse equilibria are the theoretical modelling of Sollich and collaborators, based on the moment free energy method \cite{Fasolo2004, Sollich2011}. A polydisperse equilibrium phase diagram is not presently available for the square well interaction considered here. Still, the hard sphere calculations in Ref.~\cite{Fasolo2004} yield important qualitative insights. Within the hard sphere crystal-fluid coexistence region, even small parent polydispersity is associated with reduced polydispersity in the crystal, as observed in our dynamical simulations. For higher parent polydispersity and volume fraction, there is the possibility at equilibrium of \textit{multiple} crystalline phases. However, the qualitative picture within the crystal-fluid coexistence region (so that the crystal can reject particles into the coexisting fluid), at the polydispersities we consider here, is of a fluid coexisting with a single crystal phase, of reduced polydispersity relative to the parent \cite{Fasolo2004}. Additionally, when such multiple crystal phases do enter into the phase diagram, they may not be kinetically attainable on experimental timescales \cite{Liddle2011} -- hence the question of how a single crystalline phase grows would remain important.

We note also that, within the moment free energy method, the present best assumption is that the equilibrium crystal(s) have substitutionally-disordered FCC structure, as is assumed for our crystal template. Against the backdrop of overall substitutional disorder (i.e.~an FCC-like structure, as opposed to an alloy), we found (Section~\ref{sec:localcorr}) that some local ordering is detectable, i.e.\ the crystal `notices' particle size in deciding the local distribution of particles. For high polydispersity, it has been noted \cite{Sollich2011} that substitutionally-ordered alloy structures such as $\textrm{AB}_{2}$ may become preferable. No calculations have yet dealt with this question, however. Therefore, for the present degree of polydispersity, with the best current equilibrium information, our templating strategy seems realistic, in that it allows the expected equilibrium crystal to form. 

\subsection{\label{sec:summ}Summary}

The results above show that the crystal phase is able to reduce its polydispersity relative to the fluid even as it is growing. In addition, we have measured a local radial size correlation function $\xi(r)$ to quantify the nature of local size ordering in each phase. Both the overall form of $\xi(r)$ and its relationship to $g(r)$ must change qualitatively between the fluid and crystal phases. This indicates a further, local ordering process \textit{in addition to} the overall preference of the crystal to reduce its polydispersity. To our knowledge, this phenomenon has not previously been observed, and gives new insight into the detailed structural features of polydisperse phases. We have also found that local size segregation, previously found in a polydisperse fluid near a wall, seems to have a close analogue in the structuring of a bulk polydisperse fluid.

We have proposed that the necessary fractionation and rearrangement processes must take place at the crystal interface and therefore, relying on self diffusion near the interface, are enhanced by the presence of the gas layer. If true, this mechanism would explain the very slow crystallisation in the $\sigma = 0.06$ gas-free case, where no such layer exists.

The $\sigma = 0.03$ simulations provide an interesting intermediate case, in which the crystal growth in the gas-mediated case is initially faster before being overtaken by the gas-free case. In light of our hypothesis, this may be because the gas-free growth is relatively slower until the interfacial fluid is sufficiently depleted to allow adequate diffusion near the interface. In the gas-mediated growth, a gas layer is formed \textit{immediately}, so that growth is initially faster. However, once the gas-free system has sufficiently depleted the interfacial fluid, the interfacial diffusion is enhanced and it now grows the fastest, the gas-mediated system lagging behind for the reasons discussed in Section~\ref{sec:growthrate}. In contrast, for $\sigma = 0.06$ the gas-free growth is so slow (presumably due to the greater fractionation required) that the interfacial fluid is not depleted enough for such a crossover to happen on the simulated timescale.

We stress that the ideas outlined in this section constitute only one potential explanation of the results we observed in Section~\ref{sec:results}. Further work will be required to determine other factors (e.g.\ a quantitative calculation of polydispersity-induced changes to the equilibrium phase diagram of square well particles), and to more rigorously test the dynamical explanation outlined here, which we expect will remain an important part of any later explanation. Nonetheless it is clear that, whatever the mechanisms, the influence of metastability on crystal growth kinetics is far from trivial, and that its role can be qualitatively switched by the presence of a very mild degree of polydispersity. 



\section{\label{sec:conclusions}Conclusions}

We have studied the kinetics of crystal growth using a model system in which a single crystal grows from an amorphous fluid. By varying the interaction strength and the polydispersity, we were able to investigate the separate and combined effects of metastable gas-liquid separation and polydispersity on the crystal growth process. Taking advantage of the pre-determined growth direction of the crystal, we have used one-dimensional concentration profiles to clearly display the phase ordering scenarios when the system is inside or outside the metastable G-L binodal. Inside the binodal, we observed the formation of a gaseous layer coating the crystal, in agreement with approximate free energy curves and with previous experimental observations of three-phase ordering. The diffusive growth of the crystal-gas-liquid `split interface' was modelled theoretically, and we found approximate agreement between theoretical and simulation interface growth profiles. This agreement validated our use of two approximations in the theory, namely that the interfaces are locally equilibrated to the densities given by the phase diagram (FIG.~\ref{LGK}) and that particle transport through the gas layer is approximately ideal gas-like due to the gas's low density.
	
The dynamic influence of this gaseous layer, in the monodisperse case, was to impede growth by slowing particle transport to the crystal interface, suggesting that, while metastable G-L separation can enhance crystal \textit{nucleation} \cite{Anderson2002, Fortini2008}, it may (in the ideally monodisperse case at least) have the opposite effect on the subsequent \textit{growth} of the crystal if the free energy landscape is such that the metastable liquid cannot coexist locally with the crystal, as was the case here. However, introducing a small amount of polydispersity, corresponding roughly to that found in a `near-monodisperse' colloidal system, we found that the gaseous layer instead strongly \textit{enhanced} growth. 

The crystals grown show an overall reduction in polydispersity versus the parent, consistent with existing equilibrium theory and equilibrium simulations. This fractionation is often proposed as an explanation for slow crystallisation in polydisperse systems, but to our knowledge has not previously been measured in the dynamical crystal growth of polydisperse spherical colloids. We have postulated that this fractionation process, requiring self diffusion, is facilitated by the gas layer, perhaps explaining why this layer enhances crystal growth in the polydisperse case, rather than impeding it as in the monodisperse case.

Additionally, we postulated and observed in detail a process of \textit{local} size ordering in crystal formation. We defined and measured a radial size correlation function $\xi(r)$, whose appearance resembles the radial distribution function $g(r)$ in each phase. We have described and explained the relationship between these two functions in both the crystal and fluid phases, showing that both the form of $\xi(r)$ and its relationship to $g(r)$ must be qualitatively altered as the crystal is grown. The findings for the fluid phase seem to indicate that local size segregation previously observed near a hard wall also appears in the structuring of a bulk fluid. 

These measurements indicate a local ordering process, in addition to the overall preference of the crystal to reduce its polydispersity relative to the fluid, which would also be enhanced by the presence of the gas layer. Aside from this, further study of the detailed form of $\xi(r)$, the dependence on system parameters, equilibration, etc.\ could provide a useful new angle on the influence of polydispersity on local crystal structure. For instance, preliminary work in this direction (not shown here) shows that such correlations persist even when the parent polydispersity is so low that a dynamically-grown crystal does not reduce its polydispersity \textit{at all} relative to the parent, i.e.\ overall fractionation does not take place.

It must be stressed that our proposed explanation requires further investigation and probably does not constitute the full story. Other factors, discussed in Sections~\ref{sec:eq} and \ref{sec:summ}, must be considered -- further work will be required to establish their influence, if any, and to rigorously test the explanation we outlined here. We used two state points (in $u$ and $\phi_p$) in order to focus on the effects of multiple polydispersities at these points -- it would be beneficial in future work to explore the phase diagram further, for instance by `exiting' the gas-liquid binodal in another direction (e.g.\ increasing/decreasing $\phi_p$ as well as varying $T_{\textrm{eff}}$). We have, though, mentioned simulations run at lower $T_{\textrm{eff}}$, further from the gas-liquid critical point, which suggest that the effect of the gas layer here is not dependent on critical phenomena. Also, as with other comparable simulations, hydrodynamic interactions (HI) have been neglected; more complex and expensive simulations would be required to quantify possible effects of HI on the dynamics observed here. 

We have presented a detailed dynamical study of crystal growth scenarios in the presence of metastability and polydispersity, and shown how our findings relate to the free energy landscape of the system. Whatever emerges from future work, the results demonstrate the importance of both metastability and polydispersity for soft matter phase transition kinetics and, moreover, that these two factors interact in a complex and previously unknown manner, the causes of which remain to be investigated further.

Thanks are due to P.~Bartlett and C.~P.~Royall for their interest and suggestions, and in particular to N.~B.~Wilding for a useful discussion regarding local size ordering. J.~J.~W. was supported by an Engineering and Physical Sciences Research Council Doctoral Training Grants award.

\bibliography{Main}

\end{document}